# Influence of Interplanetary Coronal Mass Ejections on Terrestrial Particle Fluxes Through Magnetosphere Disturbances

*A.Chilingarian, T.Karapetyan, B.Sargsyan, K.Asatryan, G.Gabaryan*

*Yerevan Physics Institute*
*Alikhanyan Brothers 2, Yerevan, Armenia AM0036*

**Abstract**

This study investigates the modulation of particle fluxes at the Earth's surface influenced by the intensity and orientation of the Interplanetary magnetic field (IMF) carried by the Coronal Mass Ejecta (ICME). We examine how IMF and its Bz component, opposing the magnetosphere, significantly enhance geomagnetic activity through magnetic reconnection. This reconnection facilitates increased penetration of solar wind particles into the magnetosphere, thus amplifying the fluxes registered by terrestrial particle detectors and enhancing particle fluxes through reduced cutoff rigidity (magnetospheric effect, ME). Conversely, the orientation of the Bz component is less crucial for a Forbush decrease (FD); instead, the strength of the ejecta's scalar magnetic field (B) predominates, potentially triggering a significant FD. The study explores how magnetic field variations influence the flux of neutrons and muons, effectively modifying the observed rates of cosmic ray influx. Comprehensive data from the WIND magnetometer and Aragats spectrometers underline the direct relationship between ICME magnetic configurations and variations in ground-level particle fluxes.
Moreover, we discover that the energy spectra of additional particles during ME are limited to 10 MeV due to the low energy of solar protons entering the terrestrial atmosphere. In contrast, the energy spectra of the "missing" FD particles can extend up to 100 MeV, demonstrating that magnetic traps and cradles formed by interactions between ejecta and Earth's magnetic fields can also deflect medium-energy solar protons. These insights advance our understanding of geomagnetic modulation of particle fluxes and bolster predictive models of space weather impacts on particle detection technologies.

1. Introduction

The relationship between the fluxes of neutral and charged particles measured on the Earth's surface and geomagnetic storms (GMS) caused by approaching Interplanetary Coronal Mass Ejections (ICME) is currently the subject of intense research, especially with the upcoming maximum of the 25th solar activity cycle. This dynamic interplay induces variations in the typically stable influx of galactic cosmic rays into the terrestrial atmosphere. Cosmic rays are monitored on Earth by networks of particle detectors, which measure secondary cosmic rays produced when primary protons and stripped nuclei interact with atmospheric atoms. These networks include Neutron Monitors (NMs, Mavromichalaki et al., 2011), which register the flux of secondary neutrons, and SEVAN detectors (Chilingarian et al., 2009a), which additionally capture electron, gamma ray, and muon fluxes. The network of NMs covered almost the whole globe, from the Antarctic to the Arctic regions. SEVAN network detectors are located at mountain tops in Armenia, Germany, and Eastern Europe. The detection of solar events is surprisingly coherent by both networks (Chilingarian et al., 2024a). The secondary particle fluxes can enhance for a few hours (magnetospheric effect, ME, Chilingarian et al., 2024b) or deplete, with the following recovery lasting 1-2 days or more (Forbush decrease, FD, Forbush, 1954). The ME is explained by the decrease of the cutoff rigidity (the minimal energy allowing galactic protons to enter the atmosphere above the detector), FD – by the emerging difficulties of entering galactic protons into the atmosphere due to magnetic traps that can deflect charged particles, impacting their entry into the atmosphere. There are no exhausting models of how these two cases lead to opposite effects in secondary particle flux intensity and, overall, in understanding the origination of FDs and MEs. The relative contribution of the shock-sheath and magnetic cloud regions to FDs has generally been inconclusive. Only statistical investigations can not encompass all effects affecting the count rate of surface detectors, including event anisotropies, long and short-term periodicities induced by the solar activity cycles, conditions of the interplanetary magnetic field and magnetosphere, and others. Adding to the measured variations in the cosmic ray intensities, the energy spectra of particles reaching the detector (ME) or the deficit of counts (FD) will help to develop models of particle interaction within the disturbed magnetosphere. This will enable simulations of the transport of secondary particles in the atmosphere, allowing for a comparison between the simulated and measured energy spectra.

Magnetospheric events, directly linked to GMS, can be complicated by the simultaneous registration of secondary particles from Solar Energetic Particle (SEP) events accelerated during intense flares. SEPs contain protons energetic enough to generate secondary particles in the atmosphere, leading to a brief (several tens of minutes) increase in detector count rates, known as ground-level enhancements (GLEs). GLEs typically reach detectors approximately $\approx$ 10 minutes after a solar flare, while the arrival time of an ICME after the solar burst is much longer, usually 20 hours or more. However, the same group of solar spots can trigger several ICMEs before and after the SEP event.

All these effects depend on the detector location site's altitude, longitude, and latitude. Moreover, several FDs are anisotropic, i.e., well pronounced above several continents and weaker at others. Recently, using a modernized SEVAN detector, we added to the research on the FD, ME, and GLE the energy release estimation of particles hitting the detectors during these effects. The energy spectra of particles reaching the SEVAN detector during violent solar events allow a direct comparison of the particle flux energy spectrum with what is expected under different hypotheses on galactic proton propagation in the disturbed magnetosphere. This work provides new insights into the modulation of cosmic rays by transient solar events and contributes to our understanding of space weather's potential hazards. The updated SEVAN detector measurements offer a unique perspective on the energy spectra of particles during these events, facilitating improved predictions of solar storms' impacts on satellite operations, pipelines, and electrical grids. Our comprehensive analysis of the November 2023 and May 2024 events underscores the importance of advanced measurement techniques in studying solar-induced phenomena.

2. **Instrumentation**

SEVAN (space environment viewing and analysis network) represents a network of particle detectors strategically located at middle to low latitudes, primarily on mountain peaks (ChilingarianChilingarian et al., 2018). The idea of the Sevan network can be followed by Lev Doman's design to put muon detectors above and beneath the neutron monitor, realized by Khacik Babayan and Karlen Matevosyan in Nor Amberd multidirectional muon monitor (NAMMM, Chilingarian et al., 2009b) in 70ths of last century. The further development of the stacked particle detector measuring different species of cosmic rays was connected with the worldwide network of solar neutron telescopes (SNT) coordinated by Yasushi Myraki in the late 90ths (Muraki et al., 1995). The Aragats SNT (ASNT, Chilingarian et al., 2022) was designed by Armenian physicists and already included the base function of particle energy recovery by energy releases in thick plastic scintillators and precedes compact, although multifunctional modules of the SEVAN network. The International Heliophysical Year (IHY - 2007) initiated the distribution of the SEVAN detectors. Over nearly 15 years, SEVAN detectors have measured time series of charged and neutral particle count rates. SEVAN detectors cover various latitudes and altitudes, operating in Armenia, Croatia, Bulgaria, Slovakia, the Czech Republic, and Germany (Chilingarian et al., 2021). Notable sites include Mt. Aragats in Armenia, Mt. Musala in Bulgaria, and Zugspitze in Germany. The SEVAN modules contain three layers of particle detectors interlayered by lead absorbers and plastic moderators to be compatible with neutron monitors. However, after recognizing that muon fluxes are as sensible to Solar events as neutron ones, we shorten detectors to 2 layers without lead (see Fig. 1). The newly developed electronics board has added a new capability to these detectors, located at Aragats and Zugspitze (Chilingarian et al., 2024c). It can now measure the energy spectra of particle fluxes enriched by charged and neutral particles. The SEVAN-light detector, using the veto signal from the upper thin scintillator, which measures fluxes of electrons and muons, the signals and energy releases of neutrons and gamma rays in the lower thick scintillator can be separated neutral fluxes ("01" coincidence of the scintillators) of charged ones. The "11" coincidence selects the charged content of cosmic rays. Thus, the coincidence techniques can select neutrons from the violent solar bursts and neutrons and gamma rays produced in the Earth's atmosphere by the photonuclear reactions.

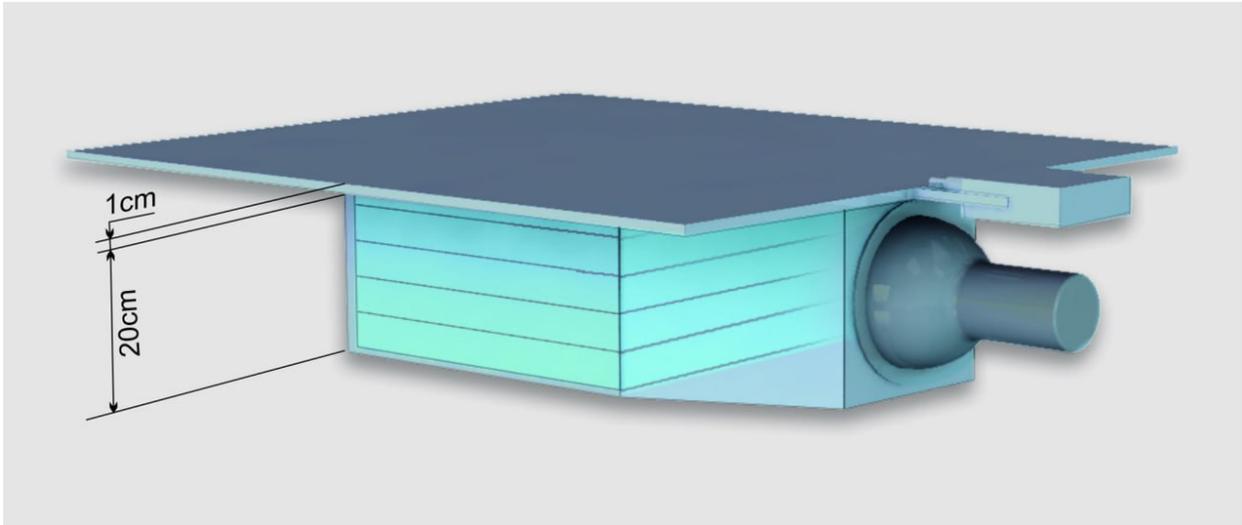

**Figure 1. Stacked detector SEVAN-light measuring charged and neutral species of cosmic rays**

The SEVAN detector, previously measuring only count rates, has been transformed into a powerful spectrometer capable of measuring energy releases of neutrons and muons in the 5-100 MeV energy range. The list of available information from modernized SEVAN will be as follows:

- 1-minute count rates of stacked five and 20-cm thick scintillators.
- 1-minute count rates of the coincidences "01", signal only in 20 cm scintillator; "10" – signal only in the upper 5-cm thick scintillator, and "11" – signal in both scintillators.
- Histograms of energy releases in both scintillators. Histograms corresponding to the coincidences mentioned above. 1-minute histograms of energy releases stored continuously.

A coincidence of "10" identifies particles that generate a signal in the upper scintillator while bypassing the lower scintillator. Thus, the particle samples and energy release histograms selected by the "10" coincidence will experience an enrichment of charged particles. In contrast, the samples and histograms selected by the "01" coincidence will be enriched by neutral particles. In addition, the "11" coincidence will select high-energy charged particles, specifically muons and TGE electrons.

Table 1 shows the results of the SEVAN light detector response calculation with simulated cosmic ray flux on Aragats obtained with the EXPACS WEB calculator (Sato, 2016). As we can see in Table 1, the "01" coincidence (signal only in the bottom scintillator) comprises ≈32% of neutrons and ≈64% of gamma rays. The contamination of charged particles is negligible; thus, with the "01" coincidence, we significantly increase the proportion of neutral particles in the sample (purity). The "11" coincidence selects ≈63% of negative and positive muons and ≈23% of electrons and positrons, and % share of neutral particles is only ≈8%. The "01" coincidence cannot differentiate between neutrons and photons. However, most of the additional neutral particles are gamma rays if we register an enhanced flux of electrons and gamma rays from the atmospheric electron accelerators (usually well above 10%). At the same time, the neutrons enter the detector during the magnetospheric effects and GLEs, not correlated with thunderstorms and usually at Aragats not exceeding 5%. Thus, TGEs (thunderstorm ground enhancements (Chilingarian et al., 2010) and solar events) can be easily distinguished.

**Table 1. The share of each of the species of cosmic ray background flux "selected" by different coincidences of the SEVAN light detector.**

| SEVAN-light | neutron | proton | mu+ | mu- | e- | e+ | gamma ray |
|---|---|---|---|---|---|---|---|
| Coin. 01 neutral | 32,1 | 0,3 | 1,0 | 0,9 | 1,0 | 0,9 | 63,8 |
| Coin.11 charged | 1,6 | 6,4 | 33,3 | 29,6 | 12,1 | 10,4 | 6,5 |

The next section will demonstrate how the newly introduced energy spectra can be applied to characterize different solar events: Forbush decreases, magnetospheric effects, and ground-level enhancements.

### 3. Energy spectra of ME and GLE particles and "missing" FD particles

The maximum of the 25th solar activity cycle is expected to occur in 2024. The peak or maximum of this cycle is anticipated to bring increased numbers of solar flares, ICMEs, GMSs, SEPs, Auroral activity, and solar particle events registered on the Earth's surface. On November 5, 2023, a significant geomagnetic storm driven by a CME interacted with Earth's magnetic field, disturbing the geomagnetic field and causing a rare magnetospheric effect (Chilingarian et al., 2024b). The G5-class geomagnetic storm that occurred on May 10-11, 2024, was one of the most intense geomagnetic events in over two decades. This storm was driven by multiple ICMEs from the sunspot cluster region 3664, which was significantly large and magnetically complex. The event was accompanied by FD, with very untypical low-intensity GLE almost uniformly distributed globally and low-intensity prolonged GLE.

In Figure 2, we present a 10-minute time series of count rates of the Nor Amberd neutron monitor (NANM) and neutron and muon fluxes selected by the coincidences of the SEVAN-lite detector. Relative time series were calculated according to the mean values measured at 19: 00-20:00. The FD amplitude was -8% for the NANM and -5% for SEVAN-light selected neutron and muon fluxes. The GLE amplitude was calculated relative to means measured at 23:00 – 24:00. It was 4% for neutrons measured by NANM and SEVAN-lite and 3% for muons measured by SEVAN-lite.

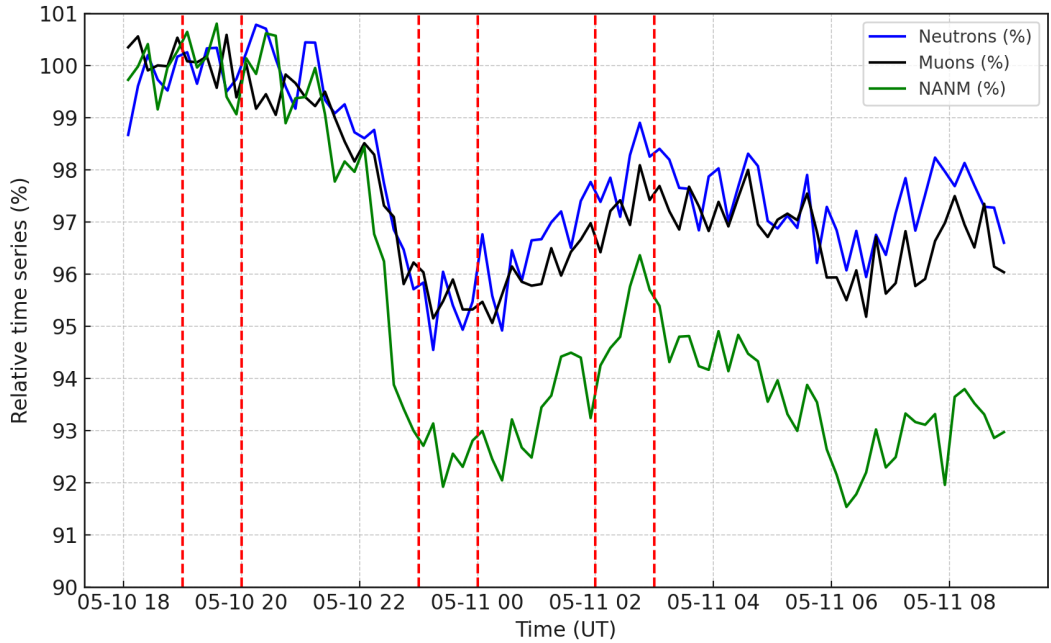

**Figure 2. Time series of count rates of Aragats and Nor Amberd detectors**

In Figure 3, we show the 1-hour energy spectra of neutrons and muons "missing" during FD (Figs. 3a and 3b) and add to the count rate during GLE (3c and 3d). To obtain the sample for FD, we subtract the counts at 23:00 – 24:00 from the counts at 19:00 – 20:00. For the GLE, we subtract the count at 2:00 – 3:00 11 May from the count of 23:00 - 24:00 on 10 May (shown by the red lines). The smooth bump in muon energy spectra is explained by the uniform energy release of charged particles in a 20 cm thick scintillator. Assuming that the energy release in each cm of scintillator is ≈2 MeV, the 40 MeV will be the most probable energy release. Neutrons should undergo nuclear reactions in the scintillator to produce charged particles, which can occur along the particle path in the scintillator. Thus, the peak in the energy release plot is smeared.

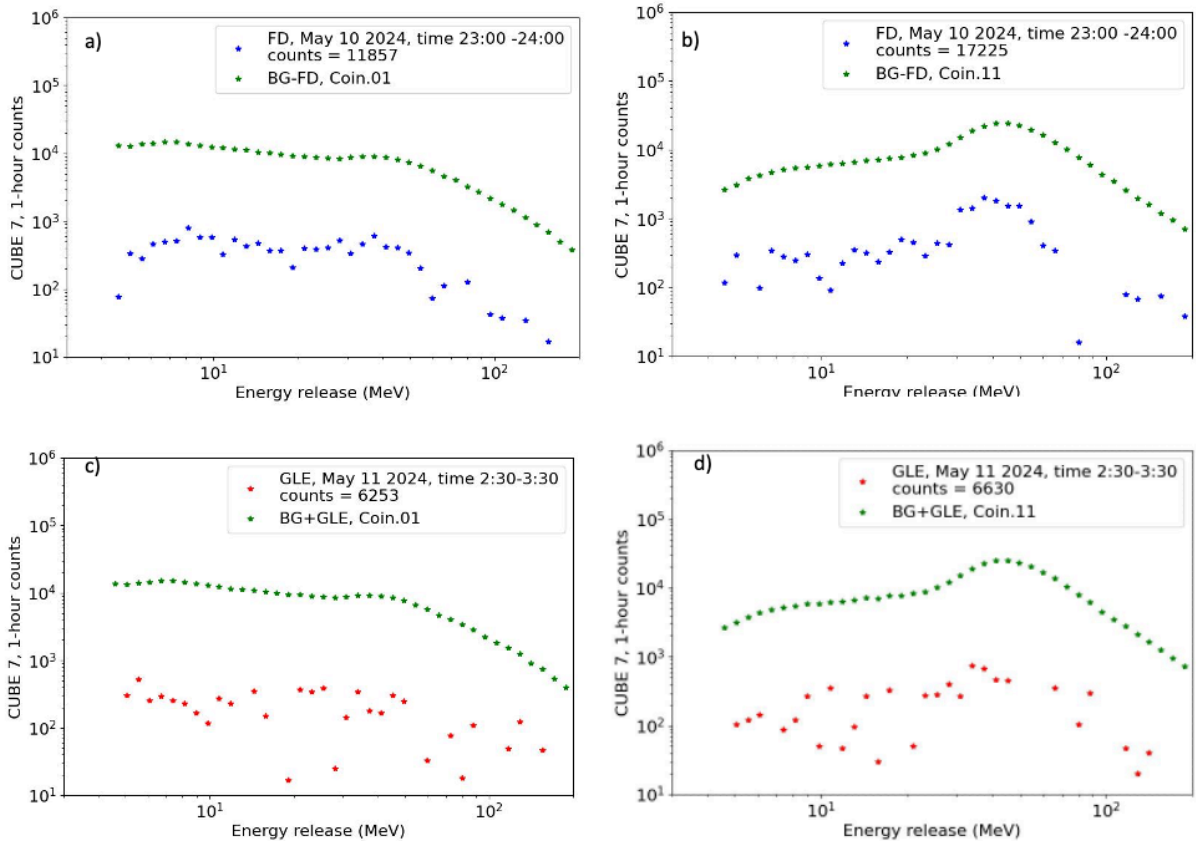

**Figure 3. The energy releases of "missing" FD particles a) and b) and of additional GLE particles measured by the SEVAN-light detector c) and d).**

As Fig. 4 shows, the energy release histogram for the magnetospheric effect strictly differed from the FD and GLE. The energy releases were measured by a 20cm thick scintillator on Aragats and 25cm thick at Zugspitze; the coincidence technique was not applied in November 2023 yet. The energies of additional particles comprising ME are mostly lower than 10 MeV. As we show in (Chilingarian et al., 2024b), it is connected that the reduction of the geomagnetic cutoff slightly below 7 GeV allows low-energy solar protons to enter the atmosphere and produce showers reaching mountain altitudes. There was no flux enhancement in Arctic/Antarctic regions and on sea level. The low-energy primary protons consequently originate low-energy secondary protons and muons.

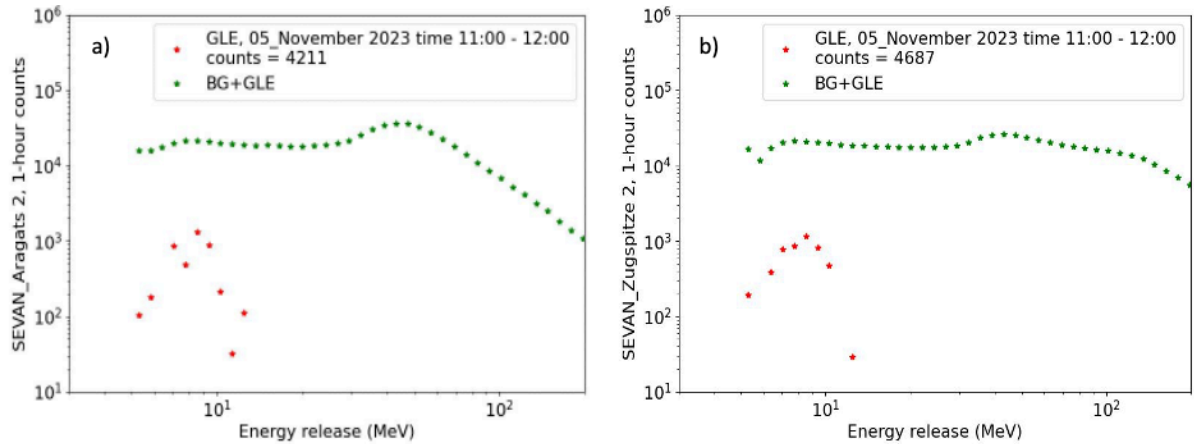

**Figure 4. The energy releases of ME particles measured by the SEVAN-light detectors on Aragats a) and at Zugspitze b).**

## 4. Discussion and Conclusions

Strong, coherent magnetic structures in the disturbed magnetosphere, which act as a barrier, reducing the influx of galactic cosmic rays, are more likely to cause significant FDs. Both southward and northward Bz components can contribute to an FD by providing magnetic shielding. A southward Bz component (negative Bz) of the ICME is crucial for geomagnetic storms. When the Bz component is southward, it is opposite to the northward direction of Earth's magnetic field. This opposition facilitates magnetic reconnection on the dayside of Earth's magnetosphere, allowing solar wind energy and particles to enter the magnetospheres. A northward Bz component (positive Bz) typically has a less severe effect on geomagnetic activity because it is aligned with Earth's magnetic field, reducing the likelihood of magnetic reconnection. ME involves an enhancement of particle flux within Earth's magnetosphere due to changes in cutoff rigidity. This is linked to suppressing the magnetosphere, lowering the cutoff rigidity, and allowing low-energy protons to enter the magnetosphere. Intensification of the solar wind speed and pressure can compress the magnetosphere and reduce the horizontal component of the geomagnetic field measured by ground magnetometers.

In Fig.5, we present measurements of the three components and total magnetic field of "cannibal" ejecta (Figs 5a-d) and the X-component of the GM measured by an Aragats magnetometer. Data on all three solar particle events are summarized in Table 2.

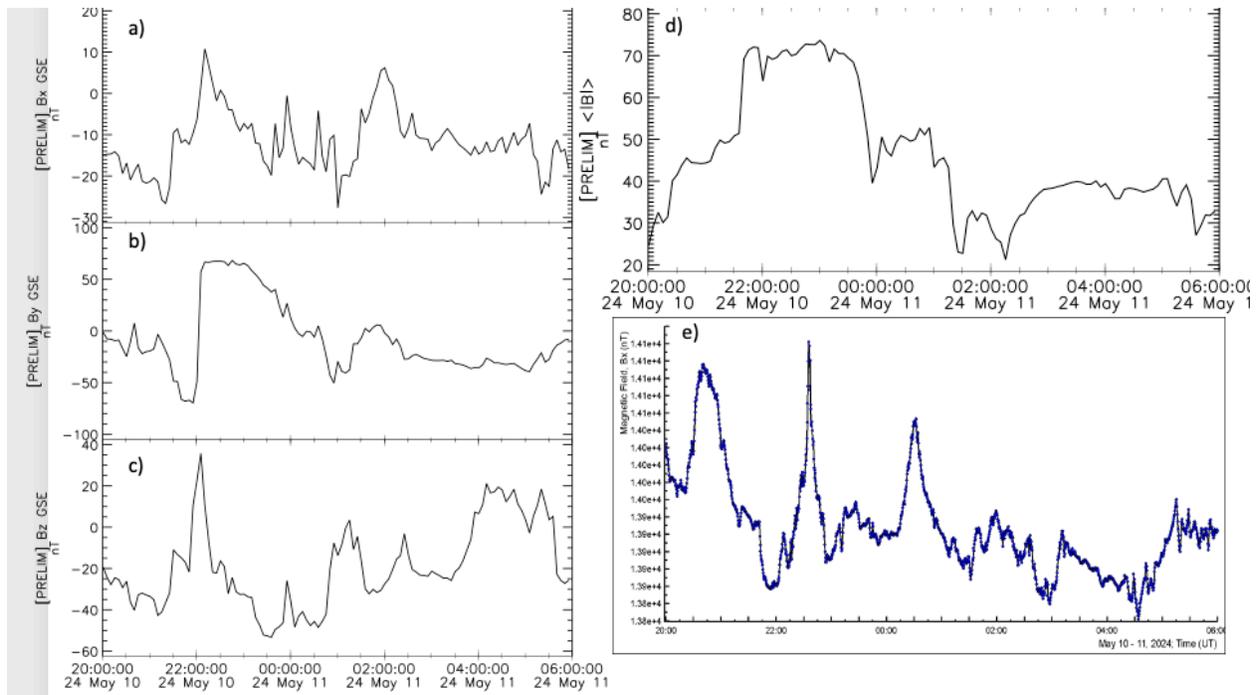

**Figure 5.** The components of ejecta magnetic field a), b), c), and scalar field d) measured by the WIND magnetometer, and the X-component of the GM field measured by the Aragats magnetometer.

In Table 2, the first two columns display the date and time of the solar particle event. The third column indicates the type of solar event, while the fourth column shows the Kp index. The fifth column presents the percentage change in particle flux. The sixth column provides the intensity of solar particle flux measured by the GOES 18 satellite detectors. The seventh column displays the values of the z component of the ejecta magnetic field measured during the event by the WIND satellite magnetometer. The eighth column shows the total magnetic field in the ejecta. The ninth column displays the Bx component of the GM field measured by the Aragats magnetometer. Lastly, the tenth column shows the maximum energy release measured by the SEVAN-light spectrometer. D sign in the Bx column notifies the oscillated GM field without any trend of reducing or enhancing.

**Table 2.** Characteristics of Solar particle events at approaching the maximum of 25ths Solar cycle.

| Date/time | Duration | Event type | Kp | % of flux chage | SEP> 1pfu | Bz  nT | Bt | Bx | Max E MeV |
|---|---|---|---|---|---|---|---|---|---|
| 5/11/2023 | 10:30-13:30 | ME | 5-6 | +5 | - | -20 - +20 | -10 - +40 | -0.5 | 10 |
| 10/5/2024 | 23:30-2:00- | FD | 8-9 | -8 | 10 | +40 | +80 | D | 100 |
| 11/5/2024 | 6:30 | GLE | 9 | +4 | 100 | -30 - 0 | +20-+40 | D | 100 |

The information from the Table confirms that the ME occurred during southward Bz and compressed GM field. FD occurred during strong GMS and northward orientation of Bz. Thus, the large absolute values of IMF are necessary for FD. The GLE, which occurred in the recovering phase of FD, didn't depend on the magnetospheric interactions; the reconnection of strong heliomagnetic fields in the corona determines it. The 11 May GLE smashed the recovery of FD, so we didn't put the end of FD of 10 May in the second column of Table 2.
The most striking feature revealed by the energy releases analysis is that strong disturbances of the magnetosphere influence particles with energies up to 100 MeV during FD and only lower than 10 MeV during ME.

## Acknowledgments

We thank our colleagues from the neutron monitor database (NMDB) and SEVAN Collaborations for engaging in valuable discussions. We are graduates of N. Ness of Bartol Research Institute and COAWeb for providing WIND data on the ICME magnetic field on 10-11 May 2024. The authors appreciate the support from the Science Committee of the Republic of Armenia, specifically for funding Research Project No. 21AG-1C012. This support was instrumental in modernizing the technical infrastructure of high-altitude stations. We also acknowledge the NMDB database (www.nmdb.eu), established under the European Union's FP7 program (contract No. 213007), for providing access to neutron monitor data.

Data availability statement: The data supporting this study's findings are available at the URL/DOI: http://adei.crd.yerphi.am/.